\documentclass[10pt,conference]{IEEEtran}
\IEEEoverridecommandlockouts
\usepackage{cite}
\usepackage{amsmath,amssymb,amsfonts}
\usepackage{algorithmic}
\usepackage{graphicx}
\usepackage{textcomp}
\usepackage{xcolor}
\usepackage{multirow}
\usepackage{subfigure}
\usepackage{tikz}
\usepackage{textcomp}
\usepackage{hyperref}
\usepackage{lipsum}
\def\BibTeX{{\rm B\kern-.05em{\sc i\kern-.025em b}\kern-.08em
    T\kern-.1667em\lower.7ex\hbox{E}\kern-.125emX}}

\DeclareRobustCommand*{\IEEEauthorrefmark}[1]{%
  \raisebox{0pt}[0pt][0pt]{\textsuperscript{\footnotesize\ensuremath{#1}}}}
  
\begin{document}
\newcommand{\umbrellaterm}{HiSEP-Q}
\newcommand\copyrighttext{%
  \footnotesize \textcopyright 2023 IEEE. Personal use of this material is permitted.
  Permission from IEEE must be obtained for all other uses, in any current or future
  media, including reprinting/republishing this material for advertising or promotional
  purposes, creating new collective works, for resale or redistribution to servers or
  lists, or reuse of any copyrighted component of this work in other works.}
\newcommand\copyrightnotice{%
\begin{tikzpicture}[remember picture,overlay]
\node[anchor=south,yshift=10pt] at (current page.south) {\fbox{\parbox{\dimexpr\textwidth-\fboxsep-\fboxrule\relax}{\copyrighttext}}};
\end{tikzpicture}%
}

\title{HiSEP-Q: A Highly Scalable and Efficient Quantum Control Processor for Superconducting Qubits
}

\author{\IEEEauthorblockN{Xiaorang Guo\IEEEauthorrefmark{1}, Kun Qin\IEEEauthorrefmark{1} and Martin Schulz\IEEEauthorrefmark{1,2}}

\IEEEauthorblockA{\IEEEauthorrefmark{1}School of Computation, Information and Technology, Technical University of Munich \\ 
\IEEEauthorrefmark{2}Leibniz Supercomputing Centre\\
Garching, Germany\\
Email: \{xiaorang.guo, kun.qin\}@tum.de, \{schulzm\}@in.tum.de}


\thanks{This work is funded by the Munich Quantum Valley~(MQV), which is supported by the Bavarian State Government with funding from the Hightech Agenda Bayern.}
}
\maketitle
\copyrightnotice
\begin{abstract}
Quantum computing promises an effective way to solve targeted problems that are classically intractable. Among them, quantum computers built with superconducting qubits are considered one of the most advanced technologies, but they suffer from short coherence times. This can get exaggerated when they are controlled directly by general-purpose host machines, which in turn get lead to the loss of quantum information. To mitigate this, we need quantum control processors (QCPs) positioned between quantum processing units (QPUs) and host machines to reduce latencies. However, existing QCPs are built on top of designs with no or inefficient scalability, requiring a large number of instructions when scaling to more qubits. In addition, interactions between current QCPs and host machines require frequent data transmissions and offline computations to obtain final results from hundreds of repeated executions, which limits the performance of quantum computers.

In this paper, we propose a QCP --- called HiSEP-Q --- featuring a novel quantum instruction set architecture (QISA)
and its microarchitecture implementation. For efficient control, we utilize mixed-type addressing modes and mixed-length instructions in HiSEP-Q, which provides an efficient way to concurrently address more than 100 qubits. Further, for efficient read-out and analysis, we develop a novel onboard accumulation and sorting unit, which eliminates the data transmission of raw data between the QCPs and host machines and enables real-time result processing. Compared to the state-of-the-art, our proposed QISA achieves at least 62\% and 28\% improvements in encoding efficiency with real and synthetic quantum circuits, respectively. We also validate the microarchitecture on a field-programmable gate array (FPGA), which exhibits low power and resource consumption, even as the number of qubits scales to 100. Both hardware and ISA evaluations demonstrate that HiSEP-Q features high scalability and efficiency toward the number of controlled qubits.
\end{abstract}


\section{Motivation}
\begin{figure}[t]
   \centering
\includegraphics[page=1,width=.49\textwidth]{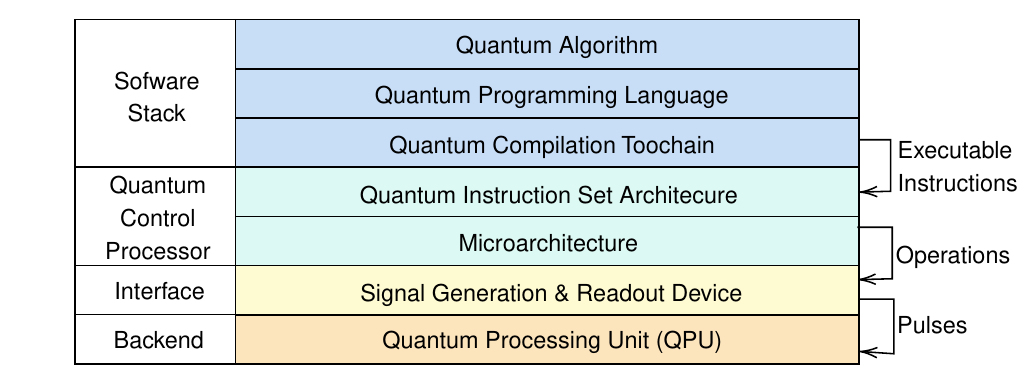}
 \caption{Abstract representation of a full-stack quantum computer.}
 \vspace{-4mm}
 \label{fig:abstraction}
\end{figure}

Quantum computing has been introduced as a revolutionary way to perform classically intractable computations and resolve complicated problems, such as cryptography~\cite{securecom2020}, molecule simulation~\cite{grimsley2019adaptive}, and optimization problems~\cite{rietsche2022quantum}. Since quantum computing is still in the Noisy Intermediate-Scale Quantum (NISQ) stage~\cite{preskill2018quantum}, developing full-stack quantum computers (shown in Fig.~\ref{fig:abstraction}) is the near-term goal to realize the full potential of its computing advantages\cite{fu2019eqasm,zhang2021exploiting}. 
Currently, research studies are primarily focusing on the software stack (upper layer of Fig.~\ref{fig:abstraction}) and hardware backend components (bottom layer of Fig.~\ref{fig:abstraction}), while the connecting layers receive comparatively less attention.
To bridge the gap, recent works have proposed designs of signal generators and readout devices~\cite{Xu2021,singhal2022generation,tholen2022measurement}. However, without control units to complement the connection, quantum computers require communications with host machines to set up the signal generation and to finish the measurement-based feedback control. The timescales of this communication easily exceed the coherence time of superconducting qubits and result in decoherence errors. This brings a huge challenge for superconducting quantum computers, though they perform better than other physical platforms in terms of gate fidelity and fabrication process.

For this reason, several groups have introduced quantum control processors (QCPs), including matching quantum instruction set architectures (QISAs) and corresponding microarchitecture designs.
These then provide real-time access for the seamless programmability and integration of the quantum systems.
QCPs reduce communications with the front-end machines and feature low-latency, real-time feedback and control of the actual quantum system. Additionally, QCPs cover the interactions with the pulse generation, which are time-critical to be able to avoid decoherence problems~\cite{zhang2021exploiting}. 

However, the existing QCPs' QISAs only support limited numbers of qubits in individual instructions, which then led to requiring many serialized instructions for concurrent qubit operations on larger quantum systems. At the same time, existing QCPs do not offer dedicated analysis units, leaving the analysis to the host system instead of offering more efficient dedicated hardware algorithms. Both issues directly limit their effectiveness on the path of large-scale quantum systems~\cite{butko2020understanding}. 

\textit{Executable Quantum Instruction Set Architecture} (eQASM)~\cite{fu2019eqasm,zhang2021exploiting,khammassi2021openql} is one of the widely adopted QISAs.
Yet, the addressing mechanism of eQASM limits not only its scalability (it only supports seven qubits), but also generality (it is topology dependent). To eliminate this drawback, Butko et al. proposed two QISAs~\cite{butko2020understanding} that achieve a larger addressing space: QUASAR and qV. However, these two ISAs still cannot provide optimal scalability, as QUASAR requires a large program size to encode quantum circuits, and qV demands massive data movements and complex hardware logic. With the increase of qubits, they will result in long execution time or large resource and power overheads, respectively.
Therefore, the limited efficiency of the current existing QISAs hinders the scalability of QCPs.

In addition to QISA challenges, due to noise interference in the NISQ era, we must repeatedly execute a quantum program hundreds or thousands of times (shots) to obtain the final result by observing the probability distribution~\cite{das2021jigsaw}. In this case, the measured results are sent back to the host machines continuously for statistical accumulation. This frequent information transmission between QCPs and host machines leads to large and unnecessary time and power consumption. 
Furthermore, when quantum computers enter the large-scale stage, the number of possible measurable states will reach $2^N$, where $N$ is the number of qubits. In this case, analyzing such a massive amount of data on the host machine is not tractable.\par

To resolve these challenges, we propose \umbrellaterm{}: \textit{A Highly Scalable and Efficient Quantum Control Processor for Superconducting Qubits}. \umbrellaterm{} includes a refined QISA based on eQASM, and the microarchitecture design for controlling superconducting qubits. Our design is able to exploit a large and efficient qubit addressing following variable length instructions combined with mixed addressing modes. Additionally, to avoid frequent transmissions and reduce the computational load of the host machine, we deploy an onboard solution for processing the results of the repeated executions of quantum programs. 

Overall, our contributions are:
\begin{itemize}
    \item We propose a mixed-type qubit addressing mechanism and a mixed-length instruction set with an optional offset indicator. These properties allow us to support a large qubit addressing space.
    \item We design an onboard histogram constructor with a top-$M$ sorting unit. 
    The occurrences of each measurement result are accumulated and sorted onboard in real-time, which reduces the data transmission overhead and provides a high-performing results analysis with negligible hardware consumption. 
    \item We evaluate our \umbrellaterm{} QISA using four benchmarks, and show that the QISA provides 62\% and 28\% improvements of program size (criteria to reflect encoding efficiency) in real and synthetic quantum circuits, respectively, compared to QUASAR. 
\end{itemize}   

We implement and validate the QCP with these properties on a field-programmable gate array (FPGA). Since the analog signal generation is not the focus of this work, the design only generates the micro-codes for the required control pulses. The FPGA is operated at 50 MHz and controlled by Python APIs. This successful validation indicates our design is capable of controlling multiple qubits with accurate time management of nanoseconds and high scalability.

\section{Background}
\label{sec:Background}
The current state-of-the-art quantum systems include a wide range of physical modalities, including superconducting qubits~\cite{schoelkopf2008wiring,Bravyi2022}, neutral atoms~\cite{Cohen21} and ion traps~\cite{Ladd2010}. Although no technology is, yet, evidently superior to others, superconducting qubits have a certain appeal regarding scalability, gate fidelity, and especially, an easier design and fabrication process with the help of conventional electronics technologies~\cite{schoelkopf2008wiring,zhang2021exploiting}. Despite the advantages, superconducting qubits require the most challenging control logic due to their short coherence time, which is the reason that our work targets this technology. Here, we provide background information on superconducting qubits and QCPs.

\subsection{Superconducting Qubits}
Superconducting qubits are operated at cryogenic temperatures to maintain their superconducting properties. Generally, we use microwave signals to manipulate qubits to control their quantum states and perform quantum operations. However, a significant challenge for superconducting qubits is their very short coherence time, which typically ranges from 50 $\mu s$ to 100 $\mu s$~\cite{Glaser23,Kjaer20}, with some superior studies exceeding 100 $\mu s$~\cite{place2021new,wang2022towards}. 
Such a short time interval makes the control logic challenging, especially with feedback and conditions, since any delays of control pulses will introduce decoherence errors and loss of quantum information.

\subsection{Quantum Control Processor (QCP)}
Due to the challenging criteria of control time for superconducting qubits, transmitting data between the host machines and QPUs for measurement-based operations is not feasible. Therefore, a control unit positioned between the software and hardware to manage control flow is a promising solution. This control unit, known as QCP, receives the binary instructions generated by the quantum compiler as input and transmits the type of gate operations to signal generators. Within the instruction stream, the state-of-the-art designs typically comprise two kinds of instructions to handle the control logic and quantum operations effectively. Quantum instructions specify the timing information and the type of quantum gate operations, while auxiliary classical instructions construct the control flow and calculate the necessary values for quantum operations. QCPs are able to separate and execute these two types of instructions while precisely controlling the timing of operations sent to hardware backends. 

\section{Related Work}
\label{sec:Related_Work}
The research on QCPs targeting superconducting qubits goes back almost to 2017, with the study by Ryan~et~al.~\cite{ryan2017hardware} being the earliest. The proposed framework features quick readout and fast feedback execution. 
Nevertheless, the complicated control and synchronization logic constrain the scalability  and parallelization of the system. Additionally, the QISA in this design lacks quantum semantics. Output waveforms are controlled by classical instructions, which require a large program size and can directly affect the processing speed of QCPs~\cite{butko2020understanding}. Similar ISA designs were proposed by multiple studies~\cite{fu2017experimental,stefanazzi2022qick}, where only simple circuits are guaranteed to meet timing requirements.

\begin{table*}[ht] 
\renewcommand\arraystretch{1.07}
\centering
\caption{Instruction set overview of \umbrellaterm{}}
\vspace{-2mm}
\label{table:QISA}
\resizebox{\textwidth}{!}{
\begin{tabular}{|llll}
\hline
\multicolumn{1}{|c|}{\textbf{Type}} &
  \multicolumn{1}{c|}{\textbf{Function}} &
  \multicolumn{1}{c|}{\textbf{Pseudoinstruction}} &
  \multicolumn{1}{c|}{\textbf{Description}} \\ \hline
  \hline
\multicolumn{4}{|c|}{\textit{original eQASM~\cite{fu2019eqasm}}} \\ \hline
\multicolumn{1}{|l|}{\multirow{6}{*}} &
  \multicolumn{1}{l|}{\multirow{2}{*}{Control}} &
  \multicolumn{1}{l|}{\textcolor{blue}{CMP} Rs,Rt} &
  \multicolumn{1}{l|}{Compare registers Rs and Rt, and store the result in the comparison flag.} \\ \cline{3-4} 
\multicolumn{1}{|l|}{} &
  \multicolumn{1}{l|}{} &
  \multicolumn{1}{l|}{\textcolor{blue}{BR} \textless{}comp.flag\textgreater{}, offset} &
  \multicolumn{1}{l|}{If the specified flag is "1", jump to address PC + offset.} \\ \cline{2-4} 
\multicolumn{1}{|l|}{} &
  \multicolumn{1}{l|}{\multirow{3}{*}{Data Transfer}} &
  \multicolumn{1}{l|}{\textcolor{blue}{FBR} \textless{}comp.flag\textgreater{},Rd} &
  \multicolumn{1}{l|}{Fetch the specified flag register to register Rd.} \\ \cline{3-4} 
\multicolumn{1}{|l|}{{Classical}} &
  \multicolumn{1}{l|}{} &
  \multicolumn{2}{l|}{\textcolor{blue}{Load \& Store} with different subtypes.} \\ \cline{3-4} 
\multicolumn{1}{|l|}{} &
  \multicolumn{1}{l|}{} &
  \multicolumn{1}{l|}{\textcolor{blue}{FMR} Rd, Qi} &
  \multicolumn{1}{l|}{Fetch the latest measurement result of qubit $i$ (Qi) into the register Rd.} \\ \cline{2-4} 
\multicolumn{1}{|l|}{} &
  \multicolumn{1}{l|}{ALU} &
  \multicolumn{1}{l|}{\textcolor{blue}{AND/OR/XOR/ADD/SUB} Rd, Rs, Rt} &
  \multicolumn{1}{l|}{Arithmetic and logical operations} \\ \hline
\multicolumn{1}{|l|}{\multirow{3}{*}} &
  \multicolumn{1}{l|}{\multirow{2}{*}{Waiting}} &
  \multicolumn{1}{l|}{\textcolor{blue}{QWAIT} Imm} &
  \multicolumn{1}{l|}{Specify a time interval (clock cycles) of waiting indicated by Imm. } \\ \cline{3-4} 
\multicolumn{1}{|l|}{{Quantum}} &
  \multicolumn{1}{l|}{} &
  \multicolumn{1}{l|}{\textcolor{blue}{QWAITR} Rs} &
  \multicolumn{1}{l|}{Specify a time interval of waiting indicated by register Rs.} \\ \cline{2-4} 
\multicolumn{1}{|l|}{} &
  \multicolumn{1}{l|}{Q.Bundle} &
  \multicolumn{1}{l|}{\begin{tabular}[c]{@{}l@{}}{[}PI, {]} Q\_Op, \textless{}target registers\textgreater{},\\ (Q\_Op, \textless{}target registers\textgreater{})\end{tabular}} &
  \multicolumn{1}{l|}{\begin{tabular}[c]{@{}l@{}}Apply gate operations (maximum two) on specified qubit targets after \\ a time interval indicated by PI (default equals 0).\end{tabular}} \\ \hline
  \hline
\multicolumn{4}{|c|}{\textit{Extended Instructions}} \\ \hline
\multicolumn{1}{|l|}{\multirow{4}{*}{Classical}} &
  \multicolumn{1}{l|}{\multirow{2}{*}{Control}} &
  \multicolumn{1}{l|}{\textcolor{blue}{\textcolor{blue}{J}} Offset} &
  \multicolumn{1}{l|}{Unconditional jump to address PC + offset.} \\ \cline{3-4} 
\multicolumn{1}{|l|}{} &
  \multicolumn{1}{l|}{} &
  \multicolumn{1}{l|}{\textcolor{blue}{END}} &
  \multicolumn{1}{l|}{Indicate the end of the program. } \\ \cline{2-4} 
\multicolumn{1}{|l|}{} &
  \multicolumn{1}{l|}{\multirow{2}{*}{Histogram}} &
  \multicolumn{1}{l|}{\textcolor{blue}{SRA}} &
  \multicolumn{1}{l|}{Start to fetch the measurement result and accumulate it in the histogram.} \\ \cline{3-4} 
\multicolumn{1}{|l|}{} &
  \multicolumn{1}{l|}{} &
  \multicolumn{1}{l|}{\textcolor{blue}{FHR} Rt} &
  \multicolumn{1}{l|}{Fetch the top $M$ results from the histogram into memory address Rt.} \\ \hline
\multicolumn{1}{|l|}{\multirow{6}{*}{Quantum}} &
  \multicolumn{1}{l|}{\multirow{3}{*}{\begin{tabular}[c]{@{}l@{}}Target Register\\ (single-qubit)\end{tabular}}} &
  \multicolumn{1}{l|}{\textcolor{blue}{SMSO} Sd, \textless{}Offset\textgreater{}, \textless{}Qubit List\textgreater{}} &
  \multicolumn{1}{l|}{\begin{tabular}[c]{@{}l@{}}Set a mask for single-qubit operations, and store it into single-qubit target \\ register Sd.\end{tabular}} \\ \cline{3-4} 
\multicolumn{1}{|l|}{} &
  \multicolumn{1}{l|}{} &
  \multicolumn{1}{l|}{\textcolor{blue}{SMSOL} Sd(l), \textless{}Offset\textgreater{}, \textless{}Qubit List\textgreater{}} &
  \multicolumn{1}{l|}{\begin{tabular}[c]{@{}l@{}}Set a long mask for single-qubit operations, and store it into single-qubit \\register Sd(l) (long instruction).\end{tabular}} \\ \cline{2-4} 
\multicolumn{1}{|l|}{} &
  \multicolumn{1}{l|}{\multirow{2}{*}{\begin{tabular}[c]{@{}l@{}}Target Register\\ (two-qubit)\end{tabular}}} &
  \multicolumn{1}{l|}{\textcolor{blue}{SITO} Td, \textless{}Offset\textgreater{},\textless{}Source\textgreater{},\textless{}Target\textgreater{}} &
  \multicolumn{1}{l|}{\begin{tabular}[c]{@{}l@{}}Set an immediate value (source and target) for two-qubit operations, and \\ store it into two-qubit register Td.\end{tabular}} \\ \cline{3-4} 
\multicolumn{1}{|l|}{} &
  \multicolumn{1}{l|}{} &
  \multicolumn{1}{l|}{\textcolor{blue}{SITOL} Td(l), \textless{}Offset\textgreater{}, \textless{}Qubit Pairs\textgreater{}} &
  \multicolumn{1}{l|}{\begin{tabular}[c]{@{}l@{}}Set an immediate value (up to seven qubit pairs) for two-qubit operations. \\ Then store the indexes into two-qubit register Td(l) (long instruction).\end{tabular}} \\ \cline{2-4} 

  
\multicolumn{1}{|l|}{} &
  \multicolumn{1}{l|}{Bit manipulation} &
  \multicolumn{1}{l|}{\textcolor{blue}{QSet} Sd/Td, \textless{}bit index\textgreater{}, 1/0} &
  \multicolumn{1}{l|}{Set the specific bit of quantum register to 1 or 0.} \\ \hline
\end{tabular}}
\vspace{-4mm}
\end{table*}

This work was followed by more comprehensive designs~\cite{fu2019eqasm,Batabyal2020,khammassi2021openql}, based on the eQASM QISA. Though this QISA can address previously mentioned problems, 
it is capable of addressing only seven qubits, and the backend support is inherently limited by the specific topology. 
Even though a modified version proposed by Zhang~et~al.~\cite{zhang2021exploiting} increases the scalability, using a superscalar structure cannot fulfill massive parallelism without single instruction multiple qubits (SIMD) support. QUASAR and qV were then proposed to enhance addressability~\cite{butko2020understanding}.
QUASAR employs a sliding mask mode to extend the addressing space beyond that of eQASM. Still, addressing a large number of qubits requires multiple instructions to complete the process.  qV supports large parallelization, but it is a purely vector-based QISA and requires significant hardware complexity. Furthermore, in both QUASAR and qV, separate timing instructions are necessary to specify individual timestamps, leading to inefficiencies. In addition, this work presents only the QISA specifications without the microarchitecture implementation and, hence, strictly seen, is not a QCP design.


\section{Novel Quantum Instruction Set}
\label{sec:QISA}
As an intermediary between the software stack and the microarchitecture, a clearly defined quantum instruction set is crucial for achieving scalable and efficient mapping of algorithms to QPUs. 
This section provides an overview of our proposed instruction set and an explanation of the new addressing mechanism. 

\subsection{Instruction Overview}
In light of the current state of quantum computing, where quantum computers serve as accelerators similar to GPUs in heterogeneous systems, it is essential to design a QISA to be compatible with classical ISAs, like RISC-V. Additionally, to effectively manage complex control flows and to ensure precise timing, the instruction set should incorporate both quantum and classical instructions and integrate explicit timing information. Therefore, we develop a quantum instruction set that builds upon the strengths of eQASM~\cite{fu2019eqasm}.

\begin{figure}[b]
\vspace{-4mm}
	\centering
    \includegraphics[width=\linewidth]{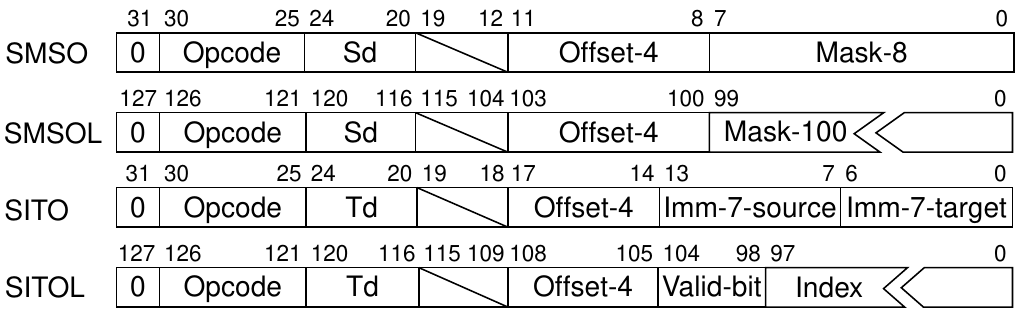}
	\caption{Instruction format for addressing target qubits. Here `Sd' (`Td) represents the target registers for single-qubit (two-qubit) operations.}
	\label{fig:address}
 \vspace{-4mm}
\end{figure}

Table~\ref{table:QISA} illustrates the overview of our proposed QISA. The top part of the table represents the original eQASM, which we use as the basic version. The bottom represents the new extended features, including novel classical and quantum instructions. The newly added features mainly bring the following benefits:


\begin{itemize}
    \item Bit manipulation (\textit{QSet}) can offer a highly effective means of encoding measurement-based feedback circuits~\cite{butko2020understanding}. Moreover, this instruction enables optimal fine-grained tuning of specific qubits, providing additional benefits for reducing the program size.
    \item To mitigate the limitation of the small addressing space in eQASM, we employ a novel mixed addressing mode (immediate and mask) and variable-length instructions with an optional offset. Within this addressing mechanism, immediate addressing provides a large addressing space and also an easier implementation. Mask addressing provides a better opportunity to realize parallelization. With the help of SIMD, long instructions can concurrently address more than 100 qubits when using single-qubit operations. For scalability, we utilize a local offset to designate the addressing range. For example, in the case of \textit{SMSOL}, if the offset filed is 0 (1), it means that the targeting is within the qubit index range of 0-99 (100-199). 
    \item Single- and two-qubit gates require a different way to encode the respective target qubits in registers. With the hurdle of distinguishing source/target operations in a mask mode, immediate addressing is more appropriate for two-qubit gates. 
    Therefore, we utilize the immediate addressing mode for two-qubit gates, which is generic and also capable of parallelization with the help of long instructions.
    \item Newly proposed histogram-related instructions (\textit{SRA} and \textit{FHR}) allow us to control the result accumulation after qubit read-out and sort the top-$M$ states from an instruction-level perspective. 
    \textit{END} instruction is used at the end of a quantum program (after repeated shots) and contributes to a more streamlined interaction within the
    quantum hardware implementation.
\end{itemize}

\begin{figure}[t]
\centering
 \includegraphics[width=.9\linewidth]{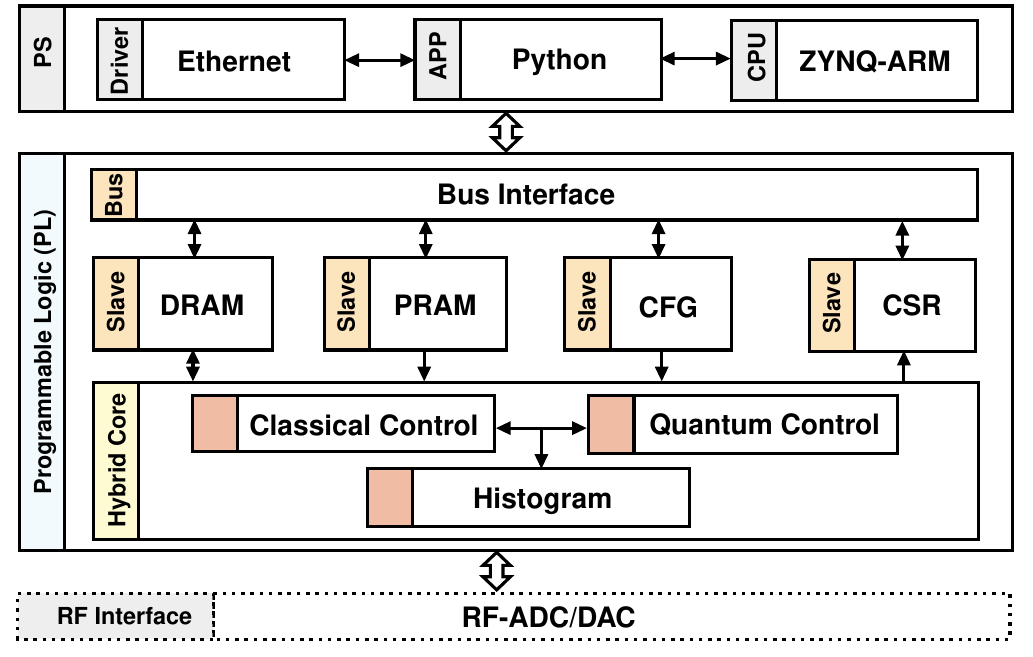}
\caption{System overview of \umbrellaterm{}}
\vspace{-4mm}
\label{fig:system}
\end{figure}

\subsection{Addressing Mechanism}
Fig.~\ref{fig:address} shows the binary encoding of our addressing mechanism. 
We utilize immediate mode for two-qubit operations and mask mode for single-qubit operations. In immediate mode, a qubit index is encoded in a 7-bit field, supporting an address space of up to 128 bits. According to the source or target operation, the two indexes are positioned in different locations, as depicted in instruction \textit{SITO}. Valid bits are employed in long instructions 
(\textit{SITOL}) to indicate if each index pair is used. \textit{SITOL} can target up to seven pairs of qubits within one instruction. In mask mode, we utilize an 8-bit mask for short instructions (\textit{SMSO}) and a 100-bit mask for long instructions (\textit{SMSOL}). 

Meanwhile, when specifying the target qubits, we use a local offset indicator within the instruction to achieve higher scalability. We assign the specified offset to each target register indicated in the respective instruction. In the current design, we use a 4-bit offset, which can theoretically address up to 1.6k qubits, but still has the potential for future extension. 


\section{Microarchitecture}
\label{sec:Microarchitecture}
In this section, we focus on the implementation of our proposed QISA in hardware. We introduce the system architecture as the starting point. Then we present the details of the structure and our proposal for the onboard histogram unit.

\subsection{System Architecture}
Fig.~\ref{fig:system} shows the schematic overview of the entire system, with a Xilinx Zynq system-on-chip (SoC) as an example. The whole system involves a processing system (PS) part and a programmable logic (PL) part. PL includes a customized AXI-bus interface, four customized slave modules, and a hybrid processing core. PS loads the program RAM (PRAM) with post-compilation instructions. 
The program execution is controlled and monitored by the \textit{configuration register} (CFG) and \textit{control and status register} (CSR).


\subsection{Hybrid Processing Core}
\begin{figure*}[ht]
	\centering    \includegraphics[width=.78\linewidth]{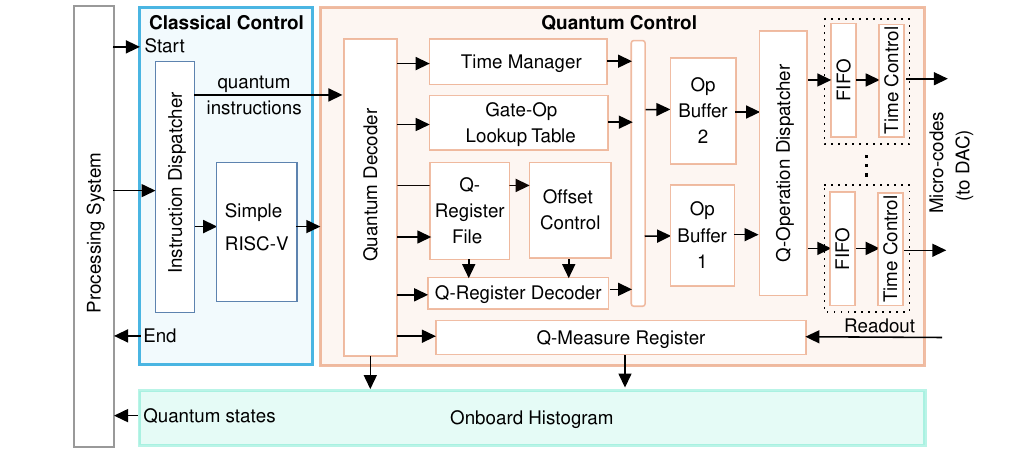}
	\caption{Microarchitecture of the hybrid core implementing \umbrellaterm{}.}
	\label{fig:hybrid_core}
 \vspace{-4mm}
\end{figure*}
As the primary implementation of \umbrellaterm{}, the hybrid core includes classical and quantum control units, and an onboard histogram unit (shown in Fig.~\ref{fig:hybrid_core}). The classical part is comparable to a simplified version of RISC-V. It is responsible for dispatching quantum instructions and executing auxiliary classical instructions for control flow and data processing. 

The quantum part is responsible for executing quantum instructions to specify the waiting time interval, target qubits, and specific operations. This part mainly contains:
\begin{itemize}
    \item \textbf{Quantum Decoder:} The quantum decoder identifies the functionalities of the incoming quantum instructions and then distributes the corresponding information (time, qubit list, or operations) to each block and generates the control signals.
    \item \textbf{Time Manager:} This block gets the timing information specified in the time-related instructions and the current timestamp from the central clock. Then it will calculate a final absolute time point for the upcoming quantum operation and put it into a timing queue.
    \item \textbf{Gate-Op Lookup Table (LUT):} Each operation encoded in the instructions should be decoded into micro-code for easier further processing. This block stores all micro-codes of gate operations and interprets the incoming instructions into the corresponding micro-codes.
    \item \textbf{Q-Register File, Offset Control \&  Register Decoder:} As there are different types of addressing modes when specifying the target registers, the length and type of registers also vary from each other. To achieve a unified output for the quantum register file, a decoder shown as \textit{Q-Register Decoder} in Fig.~\ref{fig:hybrid_core} is able to harmonize the diverse outputs and generate a single generalized format. \textit{Offset Control:} This block handles the optional offsets associated with specific registers and provides necessary information for the register decoding. The detailed mechanism will be introduced afterward.

    \item \textbf{Q-Measure Register:} This block receives the measurement results for specified qubits from read-out devices. This register only stores the latest measured result and forwards previous values to \textit{Onboard Histogram} for accumulation and sorting. 
    \item \textbf{Operation (Op) Buffer:} During the execution of each operation instruction (\textit{Q.Bundle}), three critical elements are combined to create a final operation word, which is then stored in the \textit{Op Buffer}. These components consist of the time point stored at the top of the timing queue, the specified target register, and the operation micro-code. \textit{Q.Bundle} allows instruction-level parallelism which features very long instruction word (VLIW) architectures. Since we utilize a VLIW width of 2 in the current work, we have designed two distinct pathways to fetch and process this information concurrently.
    \item \textbf{Q-Operation Dispatcher:} Based on the outputs of two \textit{Op Buffers}, this block integrates the two pathways and dispatches the micro-codes and timing information to different \textit{Timed FIFOs}, which include standard FIFOs and coupled time control units (\textit{Time Controls}), as shown in Fig.~\ref{fig:hybrid_core}. In addition, this block is also responsible for identifying whether there are multiple operations on a single qubit at the same time. If so, the program will terminate and  assert an error.
    \item \textbf{Timed FIFO:} 
    To achieve precise timing control, we adopt a queue-based methodology that ensures accurate operation sequencing. In this methodology, we assign a \textit{Time Control} to each FIFO, where one FIFO corresponds to one qubit. Once the operation dispatch is completed, operations with associated timing schedules are streamed sequentially into FIFOs. During each cycle, respective \textit{Time Controls} fetch the timestamps and compare them with the central clock. If the time points match, this block will issue operations with nanosecond-level precision according to the predetermined timetable. When current quantum operations are issued, \textit{Time Controls} fetch the timestamps from their coupled FIFOs again. 
\end{itemize}

 Among the aforementioned components, \textit{Q-Measure Register} has a specific communication directly with classical control to feature real-time feedback (we omit this path in Fig.~\ref{fig:hybrid_core} for clarity). By directly fetching the latest measurement results into the classical register file, \umbrellaterm{} can handle the measurement-based branch just at the next cycle. In addition, the onboard histogram will be introduced in detail in Section~\ref{section:histogram}.

\subsection{Quantum register decoding}
Target registers exhibit varying lengths and employ diverse index storage mechanisms. For instance, immediate values (two-qubit gates) are stored in `Td' registers, while mask values (single-qubit gates) are stored in `Sd' registers. In addition, there are special registers for long instructions. To avoid complicated dispatch logic, we adopt the method from Fu~et~al.~\cite{fu2019eqasm} and design a four-to-one register decoder. 

This decoding phase comprises two main branches: `Sd' decoding and `Td' decoding. Both branches aim to translate the register to a $2N$-bit signal, where $N$ is the number of qubits. Each qubit is associated with a two-bit indicator, specifically, `00' for no operation, `01' for source qubit, `10' for target qubit, and `11' for single-qubit operation. When translating `Sd' registers, we duplicate each bit of the mask value (replacing 0' with 00' and 1' with `11') and store the resulting two bits in their corresponding indices. While translating `Td' registers, we directly address the qubit index using the immediate value and assign the two-bit indicator to specify it as a target or source qubit. In particular, we have to consider the valid bits for \textit{SITOL}. The corresponding indicator should be set to `00' if the pair is not selected. To be noted, specific index calculations are required for registers, which are coupled with offsets.
\begin{figure}[b]
	\centering
  \vspace{-4mm}
    \includegraphics[width=.48\textwidth]{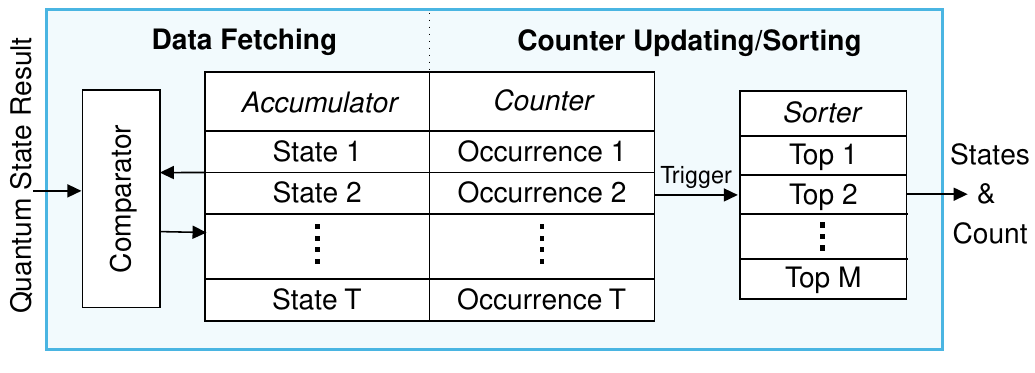}
	\caption{Architecture of the onboard histogram unit. Here T is the depth of the accumulator, which is configurable and corresponds to the number of execution shots. $M$ represents the $M$ most frequent occurrences among the quantum states.}
	\label{fig:histo}
 \vspace{-4mm}
\end{figure}

\subsection{Histogram Unit}
\label{section:histogram}
\textit{The Onboard Histogram unit} allows the accumulation of measurement results after each shot, while dynamically sorting the results on the fly. By focusing on reporting only the final quantum states with the top-$M$ probabilities, we can effectively reduce the need for frequent transmission between QCPs and host machines. Furthermore, the tremendous computational effort on the host machine, especially in large-scale quantum scenarios, can also be avoided. Consequently, this approach minimizes the communication overhead and alleviates the computational burden on classical computers. Fig.~\ref{fig:histo} illustrates our proposed histogram architecture. The depth of the histogram ($T$) is configured by the number of shots instead of possible quantum states due to the sparsity of the results. The whole process includes three separate phases.

First, when the $i$-th shot is finished, an accumulation instruction (\textit{SRA}) is issued, and data in the measurement register will be fetched to the accumulator. The newly fetched quantum state 
will be concurrently compared to the existing states stored in the accumulator. If a match occurs, the corresponding occurrence count increments; otherwise, a new accumulation branch is established, and the counter is initialized to `1'. 

Second, in the sorting phase, $M$ registers are allocated to store the top-$M$ values and their coupled quantum states. When a counter is updated (new state result is fetched and stored), we first identify if the associated state is already in the sorter. If it is within top-$M$ states, we directly update its counter in the sorter. Otherwise, we compare this counter number with the $M$-th value to determine whether the $M$-th register needs to be updated. Then the internal sorting process will start and generate the top-$M$ results after $M$+1 cycles.

In the end, when the \textit{FHR Rt} instruction executes, the top-$M$ values and their corresponding quantum states will be illustrated in the onboard histogram and stored in the DRAM for reading from the host machines.

\section{Evaluation and Experiment}
\label{sec:Experiment}
\subsection{Evaluation Setup}
\begin{figure}[t]
  \centering
  \subfigure[Grover’s operator (GO)]{\includegraphics[width=0.48\textwidth]{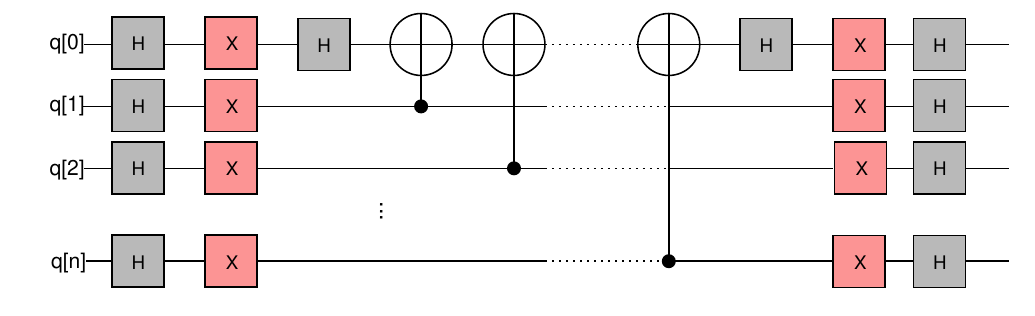}}
  \hfill
  \subfigure[Synthetic quantum circuits (gate density = 100\%)]{\includegraphics[width=0.48\textwidth]{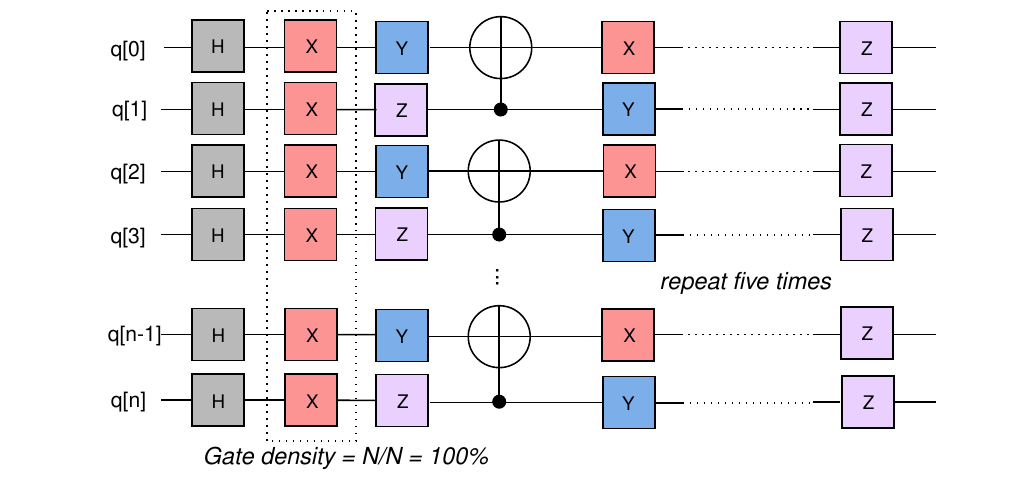}}
  \caption{Quantum circuits used as experiment benchmarks. (a) Real quantum algorithms: Grover's operator (GO). (b) Synthetic quantum circuits including four types of single-qubit gates and one type of two-qubit gate, with 100\% gate density (each qubit will be operated at each timestamp).}
  \vspace{-4mm}
  \label{fig:bench}
\end{figure}

The evaluation of our work comprises two distinct procedures: evaluating the QISA performance and validating the FPGA-based microarchitecture. We assess the QISA performance by program size, also referred to as encoding efficiency, which indicates how efficiently a quantum circuit can be encoded into a program. Smaller program size means reduced instruction memory utilization and shorter execution time. Here, we use two types of quantum circuits to evaluate the program size: a real quantum algorithm called Grover’s operator (GO)~\cite{Nielsen2011}, and a synthetic quantum circuit with different gate densities (the degree of the available gates implemented in the circuits at the same time~\cite{butko2020understanding}). The two benchmarks are depicted in Fig.~\ref{fig:bench}.

To validate the microarchitecture, we implement the \umbrellaterm{} on a Xilinx Zynq FPGA (XC7Z020-1CLG400C SoC)
and test it with several quantum programs. We focus on two key performance measures: power and resource utilization, while scaling the number of qubits. Additionally, we simulate the functionality of our implemented histogram with the online sorting unit.

\subsection{QISA Evaluation}

We compare our proposed QISA with other related works on four benchmarks: GO and synthetic quantum circuits with varied gate densities (10\% (Syn\_10), 50\% (Syn\_50), and 100\% (Syn\_100)). As these circuits are evaluated with more than seven qubits, which is not supported by the eQASM architecture, we exclude eQASM from the comparison.  

\begin{figure}[t]
	\centering
    \includegraphics[width=.4\textwidth]{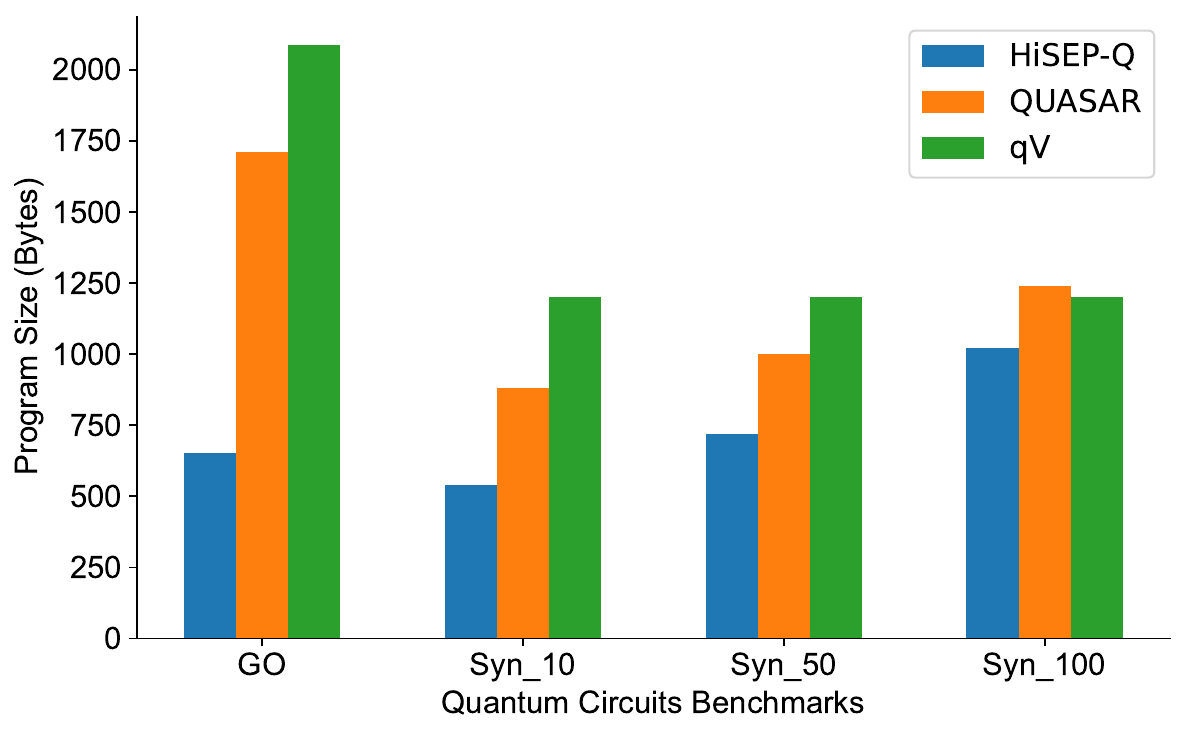}
	\caption{Program size of different QISAs under different quantum circuit benchmarks. GO represents Grover's operator algorithm, and Syn\_10 represents the synthetic circuit with 10\% gate density. Syn\_50 and Syn\_100 stand for the synthetic circuit with 50\% and 100\% gate density, respectively.}
	\label{fig:PS_QC}
 \vspace{-4mm}
\end{figure}

Fig.~\ref{fig:PS_QC} illustrates the comparison regarding the program size in different benchmarks, with the 100~qubits configuration. As depicted in the figure, our QISA demonstrates the smallest program size across all test scenarios. In particular, compared to QUASAR, \umbrellaterm{} achieves a remarkable 62\% improvement in the case of the real quantum circuit, and an average improvement of 28\% in the synthetic circuits. Due to the presence of high-density two-qubit gates in the synthetic circuits, our approach shows less improvement in synthetic circuits than in real quantum circuits.  
\begin{figure}
  \centering
  \subfigure[GO]{\includegraphics[width=0.49\linewidth]{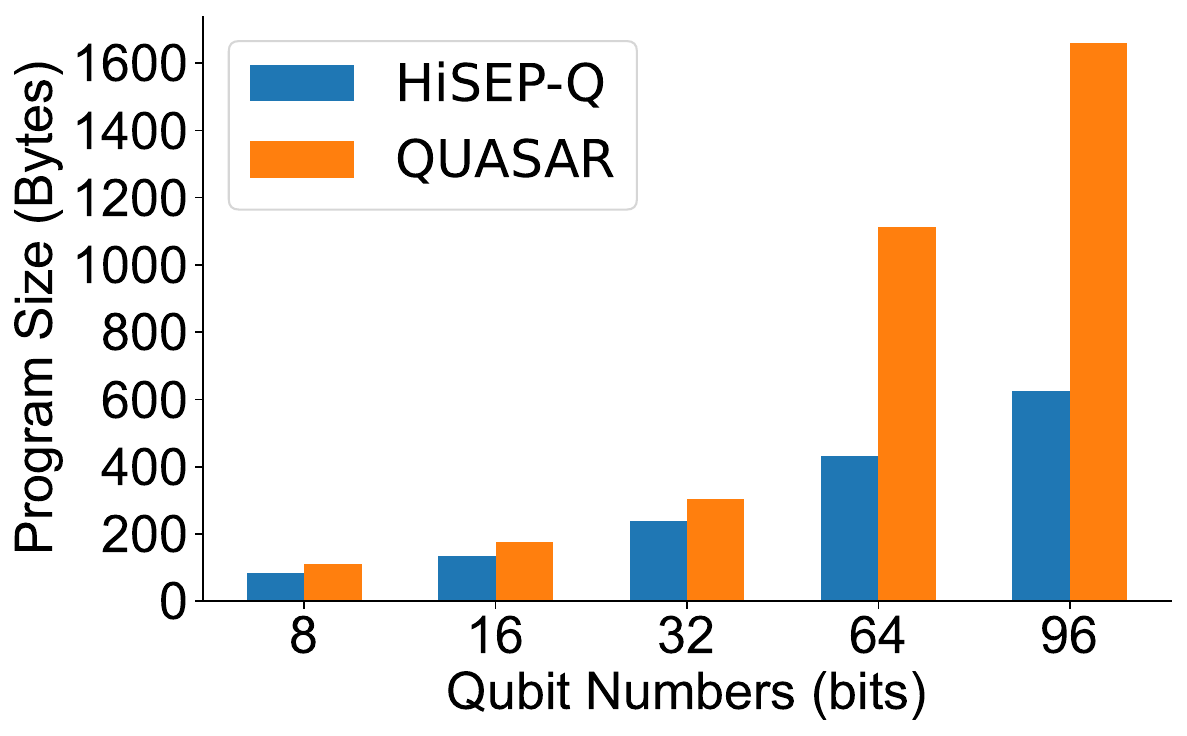}}
  \hfill
  \subfigure[Syn\_50]{\includegraphics[width=0.49\linewidth]{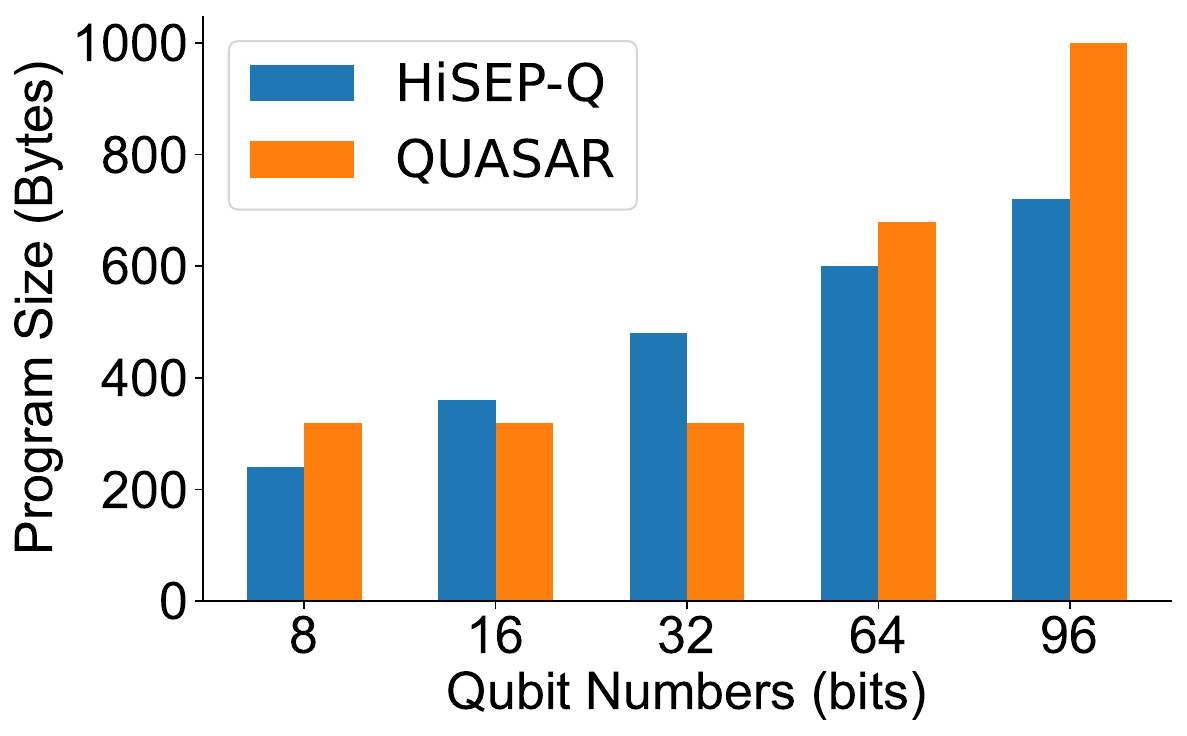}}
  \caption{Comparison of program size between \umbrellaterm{} and QUASAR when scaling the number of qubits from 8 to 96. (a) GO. (b) Synthetic quantum circuits with 50\% gate density.}
  \vspace{-4mm}
  \label{fig:PS_QN}
\end{figure} 

The program size can vary differently depending on the ISA when we scale the number of qubits, indicating the scalability of ISAs. We analyze the program size with GO and Syn\_50 using different numbers of qubits, illustrated in Fig.~\ref{fig:PS_QN}. Our comparison focuses on \umbrellaterm{} and QUASAR, which exhibits better performance than qV in Fig.~\ref{fig:PS_QC}. Both subfigures in Fig.~\ref{fig:PS_QN} consistently demonstrate that the program size of \umbrellaterm{} displays a linear increase relative to the logarithmic number of qubits, where the trend of increase is significantly lower compared to QUASAR. In Fig.~\ref{fig:PS_QN}(a), the depth of GO is proportional to the number of qubits, which explicitly increases the program size. As for Syn\_50 in Fig.~\ref{fig:PS_QC}(b), the growing qubits demand more two-qubit pairs addressed in the same timestamp, which brings more effort to the immediate addressing of \umbrellaterm{}. 
On the contrary, QUASAR shows a drastic increase in program size above 32 qubits in both evaluation cases, since the encoding becomes more consuming when the number of qubits exceeds the capacity of the mask length. Overall, \umbrellaterm{} achieves the best scalability among the related works.

In summary, there are two reasons why \umbrellaterm{} outperforms the two ISAs in these two evaluation scenarios. Using concurrent single-qubit gates as an example, QUASAR and qV require 3 and 4 pieces of 32-bit instructions, respectively, to address and manipulate 32 qubits~\cite{butko2020understanding}. When they use a sliding mask to scale up to 100~qubits, they need up to four groups of instructions (12 and 16 instructions in total, respectively). In contrast, our method only requires one long instruction and one standard instruction (equivalent to five 32-bit instructions) with a constant number of instructions when controlling less than 100~qubits. Furthermore, QUASAR and qV use specific instructions to indicate the execution time, requiring an additional timing instruction for each instruction group. Instead, this work integrates a timing interval into the \textit{Q.~Bundle}, which enables us to issue the operation and time simultaneously.

\begin{figure}
  \centering
  \subfigure[Power]{\includegraphics[width=0.33\linewidth]{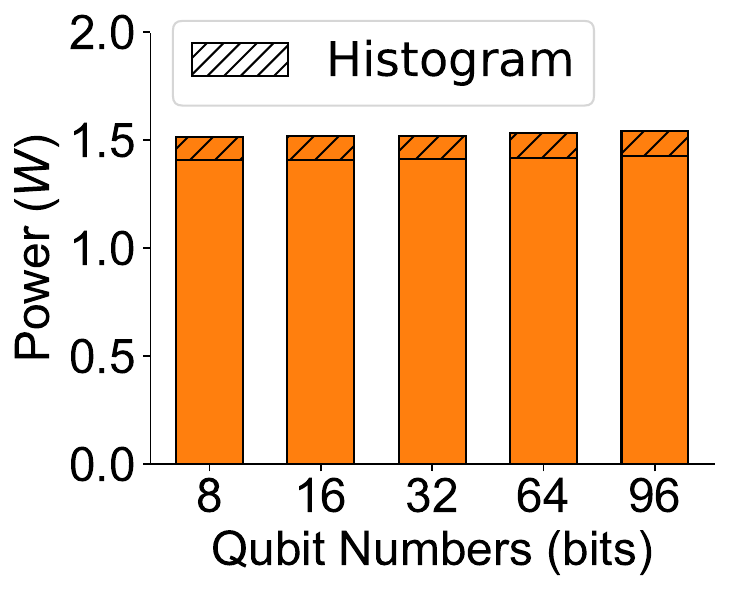}}
  \hfill
  \subfigure[Resource Utilization]{\includegraphics[width=0.62\linewidth]{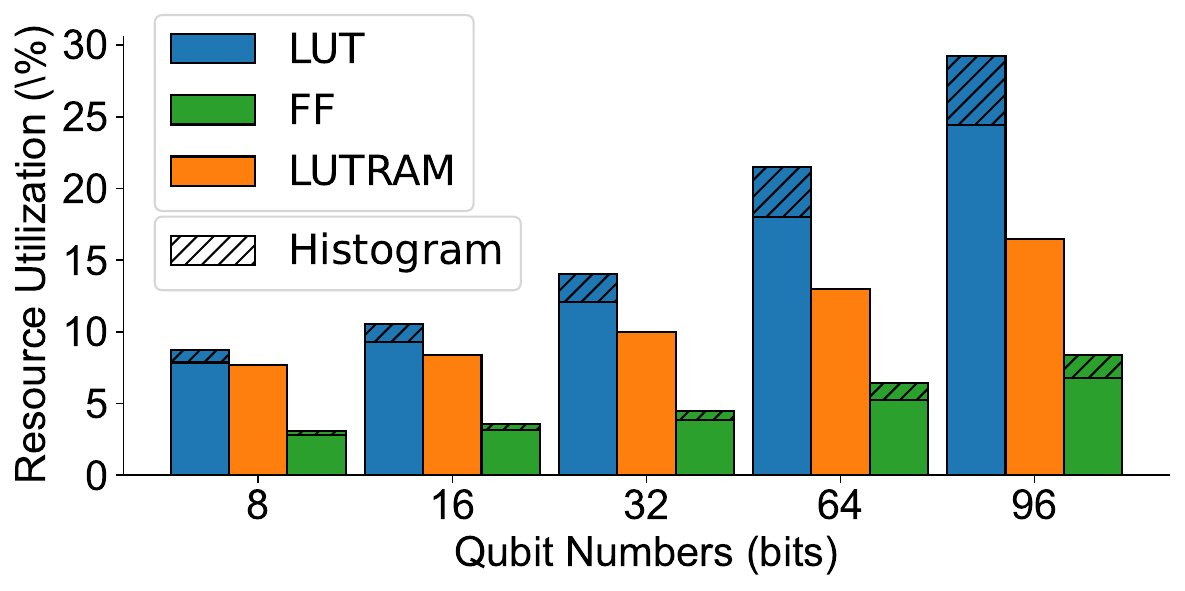}}
  \caption{Power (a) and resource utilization (b) with different numbers of controlled qubits. Crosshatched areas show the consumption of the histogram, while the non-hatched part represents the rest of the overall system. As the BRAM utilization remains the same, we do not include its values in (b).}
  \vspace{-4mm}
  \label{fig:PS_QA}
\end{figure} 

\subsection{Microarchiteture Validation}
Besides the program size, hardware performance can also restrict the scalability of QCPs. We test the hardware consumption of the overall system when it manipulates a different number of qubits, as shown in Fig.~\ref{fig:PS_QA}. To compare the resource consumption, we measure the utilization percentage of the lookup table (LUT), flip flops (FF), and distributed RAMs (LUTRAM). As the number of controlled qubits increases, the main control units and customized interfaces, which contribute to the majority of power consumption but utilize only a portion of the available resources, remain unchanged. However, the \textit{Dispatcher}, \textit{Timed FIFO}, and \textit{Histogram} components undergo reconfiguration. As a result, the power consumption of \umbrellaterm{} remains relatively stable (Fig.~\ref{fig:PS_QA}(a)), while resource utilization increases logarithmically with the number of qubits (Fig.~\ref{fig:PS_QA}(b)). It is worth noting that when we set up 96 output channels of \umbrellaterm{}, the resource utilization is only 30\% for LUT, 16\% for LUTRAM, and less than 10\% for FF, leaving enough space for integrating the signal generation logic and other extension units on the same board. Along with Fig. \ref{fig:PS_QN}, we can conclude that our proposed \umbrellaterm{} provides exceptional scalability in terms of encoding efficiency and hardware performance as the number of qubits increases.

Simultaneously, we evaluate the hardware overhead of the histogram with 100 shots of quantum circuits, whose hardware performances are crosshatched in Fig~\ref{fig:PS_QA}. The histogram only contributes to less the 10\% of the overall power, about 15\% to both LUT and FF utilizations, and no LUTRAM consumption, which is negligible for the resources on the FPGA board. In addition, we validate our onboard histogram unit with a set of Gaussian-distributed data. After feeding all the data to the histogram, we receive the top 4 results after only five cycles, which is 0.1 $\mu$s at 50 MHz. We discover that the onboard histogram can reduce in total $\frac{T-M}{T} \times N$ bit data transmission, where $T$ is the number of experiment shots, $M$ is the number of top results that we configured, and $N$ is the number of qubits. For example, when we expect the top 4 results with a 100-qubit quantum computer after running the program for 100 shots (sufficient for final result acquisition), the typical method (continuously transmitting the result to the host machines) requires 1.25k Bytes of data. In contrast, our method only needs 50 Bytes (96\% transmission reduction). Therefore, this onboard histogram can provide real-time results sorting and eliminate the transmission with host machines using relatively negligible hardware overhead.

\section{Conclusion}
\label{sec:Conclusion}
In this work, we proposed HiSEP-Q, including an efficient QISA and its architectural implementation, to support a highly scalable addressing space and onboard histogram unit. In addition, it features the crucial properties of QCPs, specifically, real-time measurement feedback and time-accurate control. In our QISA design, we employed a mixed-type addressing mode and mixed-length instruction format. Meanwhile, in the microarchitecture design, we deployed an efficient control logic, VLIW support, and an onboard histogram updating mechanism. We evaluated this work with
real and synthetic quantum circuits. The results indicate at least 62\% and 28\% improvements in encoding efficiency for real and synthetic circuits, respectively, compared to state-of-the-art works. The hardware validation of this work on a Zynq FPGA board also reveals a considerably low usage of resources and power. 
Both hardware and QISA outcomes demonstrate great scalability and encoding efficiency.
In conclusion, our work provides insights for designing highly scalable and efficient QCPs for large-scale quantum computing systems.



\bibliographystyle{IEEEtran}
\bibliography{ref}

\end{document}